# Femtosecond mega-electron-volt electron microdiffraction

X. Shen, R. K. Li, U. Lundström, T. J. Lane, A. H. Reid, S. P. Weathersby and X. J. Wang

*SLAC National Accelerator Laboratory, 2575 Sand Hill Road, Menlo Park, CA 94025, USA*

*Instruments to visualize transient structural changes of inhomogeneous materials on the nanometer scale with atomic spatial and temporal resolution are demanded to advance materials science, bioscience, and fusion sciences. One such technique is femtosecond electron microdiffraction, in which a short pulse of electrons with femtosecond-scale duration is focused into a micron-scale spot and used to obtain diffraction images to resolve ultrafast structural dynamics over localized crystalline domain. In this letter, we report the experimental demonstration of time-resolved mega-electron-volt electron microdiffraction which achieves a 5 µm root-mean-square (rms) beam size on the sample and a 100 fs rms temporal resolution. Using pulses of 10k electrons at 4.2 MeV energy with a normalized emittance 3 nm-rad, we obtained high quality diffraction from a single 10 µm paraffin ($C_{44}H_{90}$) crystal. The phonon softening mode in optical-pumped polycrystalline Bi was also time-resolved, demonstrating the temporal resolution limits of our instrument design. This new characterization capability will open many research opportunities in material and biological sciences.*

Time-resolved x-ray[1-3] and electron microbeams[4-6] are emerging tools with broad applications in science and technology. Achieving high temporal resolving power provides insight into structural dynamics of materials in non-equilibrium states for understanding and controlling of energy and matter[7]. Delivering a focused, micron-scale beam to a sample under study is of paramount importance, as it enables the study of samples that cannot be prepared at the macroscale, or are intrinsically inhomogeneous. For example, the lateral size of a "large" protein crystal is typically less than 100 µm. Further, small protein crystals almost always exhibit greater perfection and less impurity than large crystals[8]. Matching the beam size to that of the small protein crystal sample provides the best signal-to-noise (SNR) ratio for crystallography. The inhomogeneous nature of natural and nanoscale materials also requires micron-scale probes to determine local composition, chemistry, and crystalline structure. With the advent of high brightness x-ray free-electron lasers and efficient x-ray focusing optics, x-ray microbeams with femtosecond level temporal resolving power have been achieved[9-12]. Extensive research and development efforts are underway to push the frontier of electron microbeams towards the femtosecond time scales temporal resolution in the Ultrafast Electron Diffraction (UED).

Electron microdiffraction with continuous wave beams is a well-established technique in conventional transmission electron microscopy (TEM)[13]. Recently, ultrafast electron microscopy has been built by modifying a conventional TEM from thermionic or field emission electron sources to ultrafast laser-excited photoemission electron sources[14-15], allowing time-resolved optical pump – electron probe experiments. Dedicated kilo-electron-volt (keV) ultrafast electron diffraction machines have also been built and optimized[16-17]. Due to the strong space-charge forces with the moderate accelerating voltages of these instruments (200-300 keV), the beam charge density has to be reduced, sometimes to as little as a single electron per pulse, to reach a sub-ps temporal resolution while maintaining transverse emittance. Employing higher beam energy is an effective approach towards producing electron microbeams with much higher charge density while maintaining sub-ps pulse duration. Since space charge forces scale as $1/\gamma^3$, where $\gamma$ is the Lorentz factor, space-charge repulsion for an electron beam at 4.2 MeV is a factor of 200 smaller than that for a 200 keV electron beam. Therefore, it enables delivering electron bunches with orders-of-magnitude higher charge while maintaining micron-scale probe size and femtosecond-scale pulse duration. In addition, the relativistic nature of MeV electron probe naturally solves the problem of velocity mismatch in which a sub-relativistic electron probe lags an optical pump pulse, which is critical to achieve sub-100 fs temporal resolution for gas phase samples[18-20]. Electrons with higher energies also provide a larger penetration depth to enable access to thicker specimens. Therefore, the performance of UED can be tremendously improved by using electrons from high brightness MeV electron source, e.g., radio-frequency (rf) photoinjectors[21-22]. Over the past decades, the number of MeV UED instruments has been growing rapidly[23-30], which provides an excellent opportunity to achieve femtosecond electron microdiffraction.

In this Letter, we report on the experimental demonstration of femtosecond MeV electron microdiffraction from the SLAC UED instrument[30]. We present the design of the apparatus and a systematic characterization of its beam parameters. The unique advantage of micron sized probes is illustrated by capturing a single-crystal diffraction pattern of a 10 µm paraffin ($C_{44}H_{90}$) crystal, located among many closely neighbored, randomly orientated crystals. We also present the measurement results of an exemplar dynamical process, phonon softening in bismuth, which demonstrates the temporal solution of the technique.

A schematic diagram of the femtosecond MeV electron microdiffraction apparatus is shown in Fig. 1. The electron source, a photocathode rf gun and gun solenoid, is identical to that of the Linac Coherent Light Source. Counting distance from the cathode, a collimator with fixed-size apertures (100, 200, 500 µm diameter) is located at 0.56 m. A second solenoid, the micro-focusing solenoid which provides up to 0.45 T magnetic field to focus the electron beam, is located at 1.01 m. Samples are mounted at 1.36 m. Diffraction patterns are recorded on a detector at 4.60 m. A Ti:Sapphire system produces 3.2 mJ, 800 nm, 40 fs full-width-at-half-maximum (FWHM) laser that is split into two parts: a 0.6 mJ pulse is frequency tripled to 266 nm to drive the rf photocathode for electron generation, while the other 2.6 mJ pulse is used to generate pump pulses. A low level rf-laser timing system and a high stability rf power source stabilize the rms pump-probe timing jitter to 29 fs.

TABLE I. Typical operational parameters

| Parameters | | Values |
|---|---|---|
| Repetition rate | | 180 Hz |
| Beam energy | | 4.2 MeV |
| Relative beam energy spread, rms | | $7.5 \times 10^{-4}$ |
| Initial beam size on cathode, rms | | 60 μm |
| Gun phase | | 10° |
| Gun solenoid strength | | 0.25 kG-m |
| Micro-focusing solenoid strength | | 0.11 kG-m |
| Collimator diameter | | 100 μm |
| Beam charge | Before collimator | 30 fC |
| | After collimator | 1.5 fC |
| Normalized emittance after collimator | | 3.1 nm-rad |
| Probe size at the sample, rms | | 5 μm |
| Temporal resolution, rms | | 109 fs |

Our approach to generate femtosecond electron microdiffraction beams relies on three factors: (i) MeV electron beam to suppress space-charge forces to achieve sub-ps pulse duration; (ii) careful beam collimation to achieve small beam emittance; (iii) a strong solenoid lens to focus MeV electron beam. The electron beam pulse length at the sample is determined by rf compression and initial longitudinal charge density at the cathode. The gun solenoid adjusts beam divergence at the exit of the rf gun, which effectively controls the beam size at the collimator. The collimator transmits the central part of the electron beam, ensuring small divergence and high brightness. By focusing a larger beam onto the collimator aperture, an electron beam with smaller emittance but lower charge can be generated. Therefore, by changing the gun solenoid strength and collimator aperture size, it is possible to flexibly control the tradeoff between beam emittance and bunch charge. Finally, the micro-focusing solenoid determines the probe size on the sample.

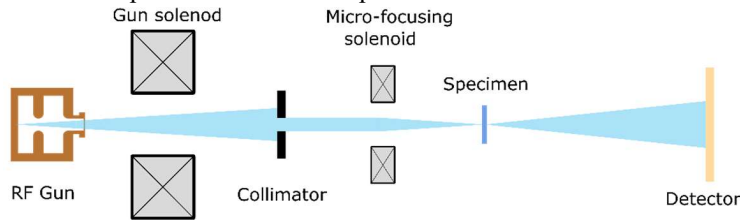

FIG. 1. A schematic diagram of the femtosecond MeV electron microdiffraction beam line.

Beam dynamics simulations using the General Particle Tracer (GPT) code[31] were conducted to guide parameter optimization. Typical machine and beam parameters are summarized in Table I. The blue and red solid curves in Fig. 2 show simulated rms horizontal beam size $\sigma_{x,rms}$ and rms pulse duration $\sigma_{t,rms}$ as a function of the distance $Z$ from the rf gun, respectively. The measured rms horizontal beam size and pulse duration at the sample position are also shown in Fig. 2 by the blue and red hollow circle, correspondingly. Details of experimental characterization of rms beam size and rms pulse duration are presented in the following sections.

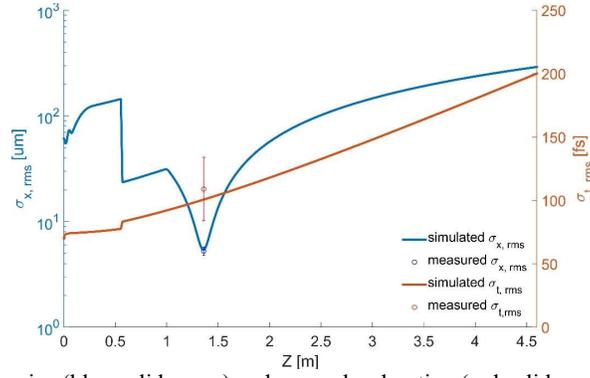

FIG. 2. Simulated rms horizontal beam size (blue solid curve) and rms pulse duration (red solid curve) as a function of distance from the rf gun. The blue and red hollow circles show the rms horizontal beam size and rms pulse duration measured at the sample position, respectively.

Characterization of beam size at the sample was conducted by the knife-edge method[32], in which a metal plate with a flat and sharp edge is scanned across the beam transverse profile. The transmitted beam intensity as a function of the sharp edge position was fitted to an error function from which the (Gaussian) rms probe size was estimated. The inset of Fig. 3 shows an example of the measured transmitted beam intensity in a vertical knife-edge scan when the beam was focused at the sample. The fitting result agrees well with the raw data and shows an estimated vertical beam size of $3.89 \pm 0.27$ μm. On average, a 5 μm minimum probe size was achieved in both horizontal and vertical directions with an optimized micro-focusing solenoid strength. The main panel of Fig. 3 shows the measured probe size as a function of the micro-focusing solenoid strength, from which the normalized transverse emittance was estimated[33] to be $3.1 \pm 1.4$ nm-rad.

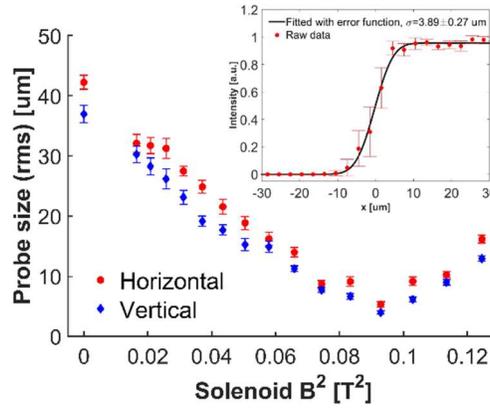

FIG. 3. Femtosecond MeV electron microdiffraction probe size characterization. Horizontal and vertical rms electron probe size at the sample were measured by knife-edge method as a function of the micro-focusing solenoid strength. The inset shows an example of raw beam transmitted intensity data and fitting result of knife-edge measurement.

The 5 μm rms beam size enables the study of localized crystalline structure. For demonstration, single-crystal diffraction patterns from paraffin ($C_{44}H_{90}$) crystals were probed with different beam sizes. Fig. 4(a) shows an optical micrograph of the paraffin sample. We took electron shadowgraph image of the sample to locate the highlighted crystal of 10 μm lateral size, and then we examined its diffraction pattern. In Fig. 4(b), the inset shows the transverse beam profile image with a 30 μm rms beam size visualized with a yttrium aluminum garnet (YAG) screen mounted on the sample holder, while the main panel shows the corresponding diffraction pattern. The almost ring-like feature exhibited in the diffraction pattern shows that multiple crystals with various orientations were sampled by the large beam. In Fig. 4(c), the inset shows the transverse beam profile image with a 5um rms beam size. The diffraction pattern in the main panel shows clear single-crystal features, although with a bigger Bragg reflection spot size compared to the case in Fig. 4(b), which indicates a lower reciprocal space resolution. At tighter focus, the beam divergence is correspondingly increased, resulting in a larger spot size in the far field. The FWHM of the Bragg spots indicates the reciprocal space resolution is 0.63 Å$^{-1}$, about four times lower than the case in Fig. 4(b). Nonetheless, the single-crystal diffraction features were well-resolved, demonstrating that the localized structural information of the 10 μm paraffin crystal was well captured by the microbeam.

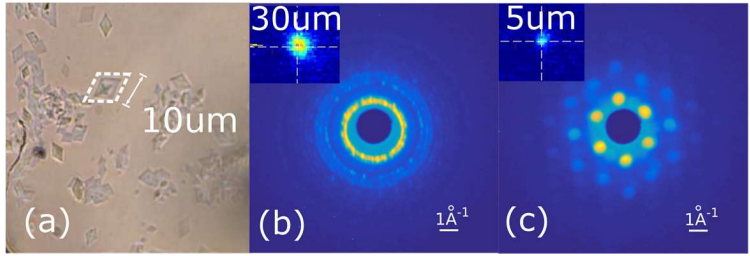

FIG. 4. Femtosecond MeV electron microdiffraction spatial resolution characterization. (a): optical micrograph of paraffin ($C_{44}H_{90}$) crystals with 10 μm size used for focus characterization. (b): diffraction pattern of the selected crystal from a 30 μm rms probe size electron beam. The transverse beam profile on a yttrium aluminum garnet (YAG) screen is shown in the inset. (c): diffraction pattern of the selected crystal from an electron beam with 5 μm rms probe size. The transverse beam profile is also shown in the inset.

For a keV electron microbeam achieved by strong focusing, the pulse duration will be stretched due to the intense space charge forces, requiring the use of a low charge density pulse to maintain temporal resolution. To demonstrate the temporal revolving power of MeV electron microbeam, we apply femtosecond MeV electron microdiffraction to measure the ultrafast structural changes of a 25-nm-thick Bi (111) thin film induced by a 60 fs 800nm pump laser. In these measurements, the micro-focusing solenoid was set on and off to generate a small and a large beam size at the sample, respectively. The inset of Fig. 5 shows a high SNR Bi diffraction pattern, which was acquired in 8 seconds with 180 Hz repetition rate. The main panel of Fig. 5 shows time-resolved measurement of the intensity of the (220) ring. The red circles represent experimental data with the micro-focusing solenoid on, while the blue squares correspond to the case when it was off. The raw data was fitted by a convolution model as $I(t) = \int_{-\infty}^{\infty} S(t')R(t-t')dt'$, where $R(t) = \frac{1}{\sqrt{2\pi}\sigma_t}\exp\left(-\frac{t^2}{2\sigma_t^2}\right)$ describes the temporal Gaussian distribution of the electron pulse with an rms pulse duration $\sigma_t$, and $S(t) = I_0$ when $t < t_0$, and $S(t) = (I_0 - I_{eq})\exp\left(-\frac{t-t_0}{\tau_{Bi}}\right) + I_{eq}$ when $t \geq t_0$, such that it represents the intrinsic intensity response of the (220) ring of the Bi (111) sample optically excited at time $t_0$ with a time constant $\tau_{Bi}$. Assuming $\tau_{Bi} = 150 \pm 50$ fs[34-36], the estimated $\sigma_t$ are $109 \pm 25$ fs and $112 \pm 29$ fs for the case with large and small probe size, respectively. These results show that beam pulse duration remains approximately 100 fs with negligible lengthening even with strong transverse focusing. These estimates set an upper bound for the electron pulse duration, since the convolution model does not consider the 30 fs rms pump pulse duration and the 50 fs rms time-of-arrival jitter between the pump pulse and the probe beam[30].

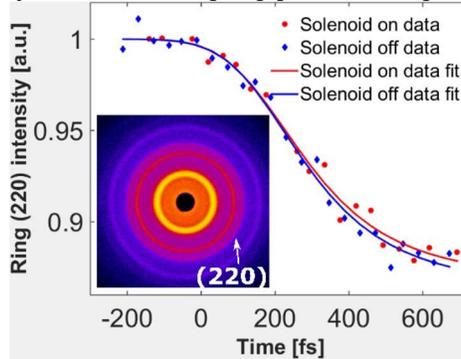

FIG. 5. Femtosecond MeV electron microdiffraction temporal resolution characterization. Intensity of the (220) ring of a 25-nm-thick Bi (111) sample after 800 nm optical excitation probed by electron beam with the micro-focusing solenoid on (red) and off (blue). Fitting the intensity change to a convolution model associated with phonon softening process and electron pulse duration yields an rms temporal resolution of $109 \pm 25$ fs for the case with solenoid off, and $112 \pm 29$ fs with solenoid on. See text for details of the fitting models.

In summary, we demonstrated a femtosecond MeV electron microdiffraction apparatus. Electron bunches at 4.2 MeV and 1.5 fC charge ($\sim 10k$ electrons) were generated and focused to 5 μm rms spot size at the sample. This beam was used to reveal the local structure information of a 10 $\mu m$ paraffin single crystal with a reciprocal space resolution of 0.63 Å$^{-1}$. The rms temporal resolution of the MeV microbeam was estimated to be 100 fs by a laser-induced dynamical process in Bi (111) thin film, and temporal resolution was not compromised by tight focusing of the beam. Further optimization of the conditions of the photocathode rf gun will help to achieve smaller beam emittance, which is essential to accomplish nano-UED[37] with sub-micron beam size and higher reciprocal space resolution. Shorter pulse duration can be simultaneously obtained by the optimized machine conditions. With the exceptional properties in beam size and temporal resolution, femtosecond MeV electron microdiffraction opens vast opportunities for study of dynamics over localized crystalline areas at atomic length and time scales in material, chemistry, and biological sciences.

The authors are grateful to their SLAC colleagues for the strong management and technical support. This work was supported in part by the U.S. Department of Energy Contract No. DE-AC02-76SF00515 and the SLAC UED/UEM Initiative Program Development Fund.